\documentclass[]{aastex631}

\shorttitle{BH activity recorded lightcurve}
\shortauthors{Zheng et al.}

\begin{document}

\title{The Activity of Black Hole Imprints on the Internal Plateau and the Subsequent Sharp Decay}



\correspondingauthor{Da-Ming Wei}
\email{dmwei@pmo.ac.cn}

\author[0000-0001-6076-9522]{Tian-Ci Zheng}
\affiliation{Key Laboratory of Dark Matter and Space Astronomy, Purple Mountain Observatory, Chinese Academy of Sciences, Nanjing 210034, China}
\affiliation{School of Astronomy and Space Science, University of Science and Technology of China, Hefei 230026, China}

\author[0000-0002-9758-5476]{Da-Ming Wei}
\affiliation{Key Laboratory of Dark Matter and Space Astronomy, Purple Mountain Observatory, Chinese Academy of Sciences, Nanjing 210034, China}
\affiliation{School of Astronomy and Space Science, University of Science and Technology of China, Hefei 230026, China}

\author[0000-0002-8385-7848]{Yun Wang}
\affiliation{Key Laboratory of Dark Matter and Space Astronomy, Purple Mountain Observatory, Chinese Academy of Sciences, Nanjing 210034, China}

\author[0000-0003-2915-7434]{Hao Zhou}
\affiliation{Key Laboratory of Dark Matter and Space Astronomy, Purple Mountain Observatory, Chinese Academy of Sciences, Nanjing 210034, China}
\affiliation{School of Astronomy and Space Science, University of Science and Technology of China, Hefei 230026, China}

\author[0000-0002-8391-5980]{Long Li}
\affiliation{School of Astronomy and Space Science, University of Science and Technology of China, Hefei 230026, China}
\affiliation{Deep Space Exploration Laboratory/Department of Astronomy, University of Science and Technology of China, Hefei 230026, China}




\begin{abstract}

A stellar-mass black hole (BH) or a millisecond magnetar is believed to be born as the central engine of Gamma-ray bursts (GRBs).
The presence of plateaus in the X-ray extended emission or afterglow of GRBs is widely accepted as an indicator of magnetar central engine, particularly those with a sharp decay (faster than $t^{-3}$), so-called internal plateau.
However, an alternative model, by taking the evolution of the magnetic flux at the BH horizon into account, suggests that an internal plateau can also arise from a Blandford-Znajek (BZ) mechanism powered jet (hereafter referred to as the BZ jet).
In this study, we propose that a precessional BZ jet would manifest a Quasi-Periodic Oscillation (QPO) signature on the internal plateau and the subsequent sharp decay.
Such lightcurves cannot be readily explained by the activity of a short-lived, supermassive magnetar, thus favoring a Kerr BH as the central engine.
The X-ray afterglow of GRB\,050904, comprising nine flares, is characterized by a QPO-modulated plateau and sharp decay, which can be well reproduced by a precessional BZ jet model.
Therefore, one potential clue for distinguishing between these two engines lies in whether QPO signature throughout the entire plateau and the subsequent sharp decay, as the magnetar scenario suggests a collapse at the end of the plateau.

\end{abstract}



\section{Introduction}
\label{sec:intro}
Gamma-ray bursts (GRBs),  known as the most energetic phenomena, release energy equivalent to the total energy of a Sun in a mere few seconds, leaving behind a mass of mist for scientists.
There are two types of central engines suggested to be born during the burst of GRBs, one being a hyper-accreting stellar mass
black hole (BH) \citep[e.g.,][]{1999ApJ...518..356P} and the other being a rapidly rotating and highly magnetized neutron star \citep[so-called magnetar; e.g.,][]{1992Natur.357..472U}.
Unfortunately, the erratic prompt emission provides scant information for identifying the type of central engine.
However, the discoveries of X-ray \citep{1997Natur.387..783C} and optical \citep{1997Natur.386..686V} afterglows offer a promising means to explore the properties of the central engine.
The canonical X-ray afterglow lightcurve \citep{2006ApJ...642..354Z,2006ApJ...642..389N} suggests that both X-ray plateaus and X-ray flares are associated with a long-lasting central engine.

A shallow decay (or plateau) followed by a normal decay is widely observed in GRB X-ray afterglows,
and can be well interpreted by continuous energy injection \citep{1998A&A...333L..87D,2006ApJ...642..354Z,2006ApJ...642..389N},
geometric effect \citep[e.g., off-set observation of a structured jet;][]{2005ApJ...626..966P},
time-dependent microphysics \citep{2006A&A...458....7I,2007ChJAA...7..509W} or environment-influenced fireball radiation \citep{2022NatCo..13.5611D}, etc.
For a plateau followed by a rapid decay, with a decay index $\alpha \sim  -2.0$, the most popular model being energy injection or the internal dissipation of a magnetar wind \citep{1998A&A...333L..87D,2001ApJ...552L..35Z}.
In some peculiar cases, the plateau followed by a sharp decay $\alpha < -3.0$ \citep[e.g.,][]{2007ApJ...665..599T,2007ApJ...670..565L}, dubbed ``internal" plateau \citep{2010MNRAS.402..705L},
where the break is associated with the disruption of the central engine, e.g., the collapse of a supermassive millisecond magnetar \citep{2010MNRAS.402..705L,2010MNRAS.409..531R,2013MNRAS.430.1061R}.
By considering the evolution of the magnetic flux at the BH horizon,
an internal plateau can be also derived from a Blandford-Znajek (BZ) mechanism \citep{1977MNRAS.179..433B} powered jet \citep{2015ApJ...804L..16K,2015MNRAS.447..327T} (hereafter referred to as the BZ jet),
though this is yet to be extensively discussed.

It has been suggested that X-ray flares and prompt emission may share a common origin, with the former potentially being induced by the intermittent accretion of a long-lasting central engine \citep{2005ApJ...630L.113K,2006ApJ...636L..29P}.
\cite{2021RAA....21..300Z} establish a connection between X-ray flares and the magnetar plateau, aiming to investigate the existence of a potential disk surrounding the central magnetar,
where the successive flares upon the magnetar plateau were suggested to have periodicity.
Quasi-Periodic Oscillation (QPO) behavior is a general phenomenon in astrophysics, observed in various systems such as active galactic nuclei \citep{1990A&A...229..424L}, X-ray binary system \citep{1980ApJ...238L.129S}, and softer gamma-ray repeater \citep{2022ApJ...931...56L}, etc.
\cite{1999ApJ...520..666P} proposed that a Lense-Thirring effect \citep{1918PhyZ...19..156L} induced jet precession would leave a pattern on GRB prompt emission. Additionally, \cite{2007A&A...468..563L} conducted a morphological study, while \cite{2023ApJ...955...98L} initiated a study to explore the corresponding manifests in sGRBs.
Recently, some potential QPO signals have been identified in both the prompt phase \citep{2022arXiv220507670Z,2022arXiv220502186X,2022ApJ...935..179W,2023RAA....23k5023Z}
and the afterglow phase \citep{2020ApJ...892L..34S,2021ApJ...921L...1Z,2022MNRAS.513L..89Z}.
Therefore, a lightcurve exhibiting QPO characteristics can serve as a valuable tool for monitoring the activity of the central engine.
In this study, we propose that the precession of a BZ jet induced by the Lense-Thirring effect will imprint a QPO signature on the internal plateau and the subsequent sharp decay.
Such lightcurves cannot be readily interpreted as the activities of a short-lived, supermassive magnetar and instead suggest the presence of a Kerr BH as the central engine.
The model successfully reproduces a distinctive X-ray lightcurve of GRB\,050904, which consists of nine flares. 
This paper is organized as follows. In \S \ref{sect:2}, we demonstrate that, by taking into account the evolution of the magnetic field and the Lense-Thirring effect, the BZ jet can generate a lightcurve, which is characterized by a QPO signature printed on the internal plateau and the subsequent sharp decay.
The model is applied to the peculiar case of GRB\,050904 and presented in \S \ref{sect:3}.
In \S \ref{sect:4}, we test the periodicity of the observed lightcurve of GRB\,050904. In \S \ref{sect:5} we give the conclusions and the discussions.
The $\Lambda$CDM cosmology parameters in a flat universe $H_0 = 67.4 ~\rm {km\ s^{-1}\ Mpc^{-1} }$ and $\Omega_M= 0.315$ \citep{2020A&A...641A...6P} are adopted in this study.

\section{emission from a precessional BZ jet}
\label{sect:2}
Currently, two models are involved in interpreting the observed internal plateau in the afterglow of GRBs.
The first model involves an internal dissipation wind originating from a supermassive magnetar, in which the break of the plateau signifies an abrupt collapse of the magnetar.
The second model pertains to an internal dissipation jet powered by BZ mechanism \citep{2015MNRAS.447..327T,2015ApJ...804L..16K}, where the break of the plateau corresponds to the expansion of the magnetosphere.
One distinguishing factor between these two models is that the latter has a continuously active central engine throughout the plateau and the subsequent sharp decay, and the QPO signature can help to record this continuous activity.

\subsection{jets powered by BHs}
\label{sect:BZJet}

The dimensionless spin parameter of a Kerr BH with mass $M_{\rm BH}$ can be expressed as $a = {\rm c}J_{\rm BH}/{\rm G}M^2_{\rm BH}$, where $J_{\rm BH}$ is the angular momentum of the BH, c and G represent the speed of light and the gravitational constant, respectively.
The magnetic flux passing through the event horizon of the BH is represented by $\Phi_{\rm BH} \sim \pi r^2_H B_H$, where $r_H$ is the radius of the outer horizon given by $r_H = {\rm G}M_{\rm BH}(1+(1-a^2)^{1/2})/c^2$, and $B_H$ represents the magnetic field strength at the horizon.
The power of the BZ jet is highly dependent on the magnetic flux, that is \citep[e.g., ][]{1977MNRAS.179..433B,2011MNRAS.418L..79T}
\begin{eqnarray}
L_{\rm BZ} \approx \frac{kf}{4\pi {\rm c}}~\Phi^2_{\rm BH}~\Omega^2_H,
\end{eqnarray}
where $k \approx 0.05$, the angular frequency of BH horizon $\Omega_H = a{\rm c}/2r_H$. For $a \sim 1$, one have $f = 1$, $r_H \sim {\rm G}M_{\rm BH}/c^2$.
The scenario with $a \sim 1$ is utilized for all calculations conducted in this study.

According to the BH no-hair theorem, the magnetic field inherited from the progenitor is expected to naturally diffuse out of the BH, as the BH formed.
The mass accretion rate of the magnetically arrested disc (MAD), however, is so high that the magnetic flux is squeezed at the BH horizon \citep{2015MNRAS.447..327T}.
The cross-section of the magnetosphere, which the magnetic flux passes through, depends on the balance between the ram pressure of the accretion flow $P_f$ and the pressure exerted by the magnetic field $P_B$.
In the scenario of magnetic flux conservation, the magnetic pressure is expressed as
\begin{eqnarray}
P_B = \frac{B^2_{\rm H,0}}{8\pi}\left(\frac{R}{r_H} \right)^{-4},
\label{eq:PB}
\end{eqnarray}
where $B_{\rm H,0}$ is the characteristic magnetic field strength while the magnetic field is squeezed at the BH horizon, one has $B_H = B_{\rm H,0} \left(\frac{R}{r_H} \right)^{-2}$, for $R \geq r_H$.
The ram pressure of accretion flow is denoted as
\begin{eqnarray}
P_f = \frac{GM_{\rm BH}\dot{M}}{2\pi R^3 v_R},
\label{eq:PF}
\end{eqnarray}
where $v_R = \epsilon (GM_{\rm BH}/R)^{1/2} $ is the radial velocity, and $\epsilon \sim 10^{-2}$ \citep[e.g.,][]{2011MNRAS.418L..79T,2014Natur.510..126Z}.
The mass accretion rate $\dot{M}$ increases with time in the early phase ($t<t_p$) and decreases with time in the later phase ($t>t_p$), where $t_p$ represents the characteristic transformation time scale of the mass accretion rate. For the later phase \citep{1989ApJ...346..847C}
\begin{eqnarray}
\dot{M} = \dot{M}_p \left( \frac{t-t_0}{t_p-t_0} \right)^{-5/3},
\label{eq:MDOT}
\end{eqnarray}
where $t_0$ is the initial time of accretion, and $\dot{M}_p$ is the characteristic mass accretion rate at $t_p$.
As the mass accretion rate $\dot{M}$ decreases and falls below a critical value of $\dot{M} \sim 10^{-4} ~\rm M_\odot~s^{-1}$ \citep{2015MNRAS.447..327T}, the ram pressure drops below the magnetic pressure level.
Consequently, the magnetosphere will move away from the BH horizon, leading to the diffusion of a portion of the magnetic flux beyond the horizon.
The magnetic flux passing through the BH horizon decreases as the magnetosphere size ($R_m$) increases
\begin{eqnarray}
\Phi_{\rm BH} =  \pi r^2_H B_{\rm H,0} \left(\frac{r_H}{R_m} \right)^{2},
\end{eqnarray}
where
\begin{eqnarray}
\frac{r_H}{R_m} =5.28 \left(\frac{\epsilon}{10^{-2}} \right)^{-2/3} \left(\frac{B_{\rm H,0}}{10^{15} ~\rm G} \right)^{-4/3} \left(\frac{M_{\rm BH}}{3~{\rm M_\odot}} \right)^{-4/3} \left(\frac{\dot{M}_p}{10^{-4}~ {\rm M_\odot} \rm s^{-1}} \right)^{2/3} \left( \frac{t-t_0}{t_p-t_0} \right)^{-10/9}.
\end{eqnarray}
Therefore, the power of the BZ jet can be written as
\begin{eqnarray}
L_{\rm BZ} &\approx & 1.44\times 10^{52} \left(\frac{\epsilon }{ 10^{-2}}\right)^{-8/3}~\left(\frac{B_{\rm H,0}}{10^{15} {\rm G}}\right)^{-10/3}~ \left(\frac{M_{\rm BH}}{3~{\rm M}_\odot}\right)^{-10/3}~\left(\frac{\dot{M}_p}{10^{-4} ~{\rm M}_\odot}\right)^{8/3} ~ \left( \frac{t-t_0}{t_p-t_0} \right)^{-40/9} ~{\rm erg~s^{-1}}.
\label{eq:L_propto_t}
\end{eqnarray}
The characteristic timescale for the diffusion of the magnetic field out of the horizon can be derived by balancing the ram pressure with the magnetic pressure
\begin{eqnarray}
t_b &=& \frac{4.47}{t_p-t_0} \left(\frac{\epsilon}{10^{-2}} \right)^{-3/5} \left(\frac{B_{\rm H,0}}{10^{15} ~\rm G} \right)^{-6/5} \left(\frac{M_{\rm BH}}{3~{\rm M_\odot}} \right)^{-6/5} \left(\frac{\dot{M}_p}{10^{-4}~ {\rm M_\odot} \rm s^{-1}} \right)^{3/5} ~{\rm s}.
\label{eq:break_T}
\end{eqnarray}
Therefore, the time-evolving power acts as a plateau followed by a sharp decay
\begin{equation}
L(t) \approx
\left\{
\begin{array}{ll}
L_0,& \quad t\leq t_{b},\\
L_{\rm 0}\left(\frac{t}{t_b}\right)^{-40/9}, & \quad t > t_{b},
\end{array}
\right.
\label{Eq:prior}
\end{equation}
where characteristic luminosity is expressed as $L_0 = \frac{\pi {\rm c}}{320 } r^2_H B^2_{\rm H,0}$. 
A smooth-broken-power-law function is employed to describe an observed internal plateau and subsequent sharp decay.
\begin{eqnarray}
L_{\rm EM} = 2\eta_{\rm BZ} L_0\left( \left(\frac{t}{t_b}\right)^{\alpha_1}+\left(\frac{t}{t_b}\right)^{\alpha_2} \right)^{-1},
\end{eqnarray}
where $\alpha_2$ represents the decay slope post the break time, with a typical value of 40/9, and the supplemented $\alpha_1$ corresponds to the decay slope before the break time, which connects the evolution of the spin of the BH and the radius of the outer horizon.
The radiative efficiency of the BZ jet is denoted by $\eta_{\rm BZ}$, with a typical value of approximately 0.1, as adopted from previous studies \citep{2015ApJ...804L..16K,2007ApJ...655..989Z}

\subsection{Precession of BZ jet}
GRBs are believed to originate from a collimated jet.
An ultra-relativistic BZ jet presents a promising possibility for yielding a beaming GRB, in the context of a BH central engine.
\cite{1999ApJ...520..666P} has proposed that the intrinsic variability in certain GRB lightcurves may arise from the precession of a jet.

In the case of a Kerr BH, its strong angular momentum exerts a dragging effect on the neighbor space.
This effect induces the inclined orbit plane of a particle to precess around the central BH,
which is known as the Lense-Thirring effect \citep{1918PhyZ...19..156L}.
Moreover, the accretion disk surrounding the Kerr BH is also influenced by this effect.
As discussed in \S~\ref{sect:BZJet}, the MAD with a high mass accretion rate would squeeze the magnetic flux on the BH horizon.
Consequently, the precessing MAD compels the direction of the BZ jet to deviate from the spin axis of the BH and to align with the axis of MAD, as described by \cite{1999ApJ...520..666P}.
Finally, the jet exhibits the same precession period as the MAD.
The precession angular frequency of the inclined plane is expressed as \citep{2013ApJ...762...98L,1972PhRvD...5..814W}
\begin{eqnarray}
\Omega = \frac{2GJ_{\rm BH}}{{\rm c}^2 r^3},
\label{Eq:Period_pressional}
\end{eqnarray}
where $r$ is the radius of the precession orbit, 
the precession period is $P = 2\pi/\Omega$.

\subsection{Evolving lightcurve originates from a precessing Jet}
\label{sect:Obpj}

Since the jet is precessing, the angle between the observer's line of sight $\vec{r}_{\rm obs}$ and
the direction of jet $\vec{r}_{\rm jet}$ should evolve with time, and it can be written as \citep[][]{2007A&A...468..563L,2010A&A...516A..16L}
\begin{eqnarray}
{\rm cos}(\psi) &=&\vec{r}_{\rm obs} \cdot \vec{r}_{\rm jet} \nonumber\\
    &=& {\rm cos}(\theta_{\rm jet}){\rm cos}(\theta_{\rm obs})+{\rm sin}(\theta_{\rm jet}){\rm sin}(\theta_{\rm obs}){\rm cos}(\phi_{\rm jet} - \phi_{\rm obs}),
\label{eq:angle}
\end{eqnarray}
where the $\theta_{\rm jet}$ and $\theta_{\rm obs}$ are the angles from the precession axis to the jet and to the observer, respectively.
$\phi_{\rm jet} = \phi_{\rm jet,0} + 2\pi t/P$ is the phase of jet, where $\phi_{\rm jet,0} $ corresponds to the initial phases,
the phases of the observer $\phi_{\rm obs}$ is a constant. By setting constant $C = \phi_{\rm jet,0}  - \phi_{\rm obs}$, a time-dependent $\psi$ is written as
\begin{eqnarray}
{\rm cos}(\psi(t)) = {\rm cos}(\theta_{\rm jet}){\rm cos}(\theta_{\rm obs})+{\rm sin}(\theta_{\rm jet}){\rm sin}(\theta_{\rm obs}){\rm cos}(2\pi t/P+C).
\label{eq:angle_p}
\end{eqnarray}
Taking into account the Doppler effect, the observed flux of an off-axis observer is written as $F(\psi) = D^3 F_{\psi =0}$,
and the Doppler factor $D = (1-\beta)/(1-\beta {\rm cos}(\psi))$, $\beta = (1-1/\Gamma^2)^{1/2}$,
where $\Gamma $ is the Lorentz factor.
Finally, the emission observed from a precessing BZ jet can be summarized into
\begin{eqnarray}
L_{\rm obs} = 2\eta_{\rm BZ} D^3 L_0\left( \left(\frac{t}{t_b}\right)^{\alpha_1}+\left(\frac{t}{t_b}\right)^{\alpha_2} \right)^{-1}.
\label{eq:LEM}
\end{eqnarray}

\section{potential signature of precessing BZ jet in X-ray flares of GRB\,050904}
\label{sect:3}
\begin{figure}[ht]
\centering
\includegraphics[angle=0,scale=0.5]{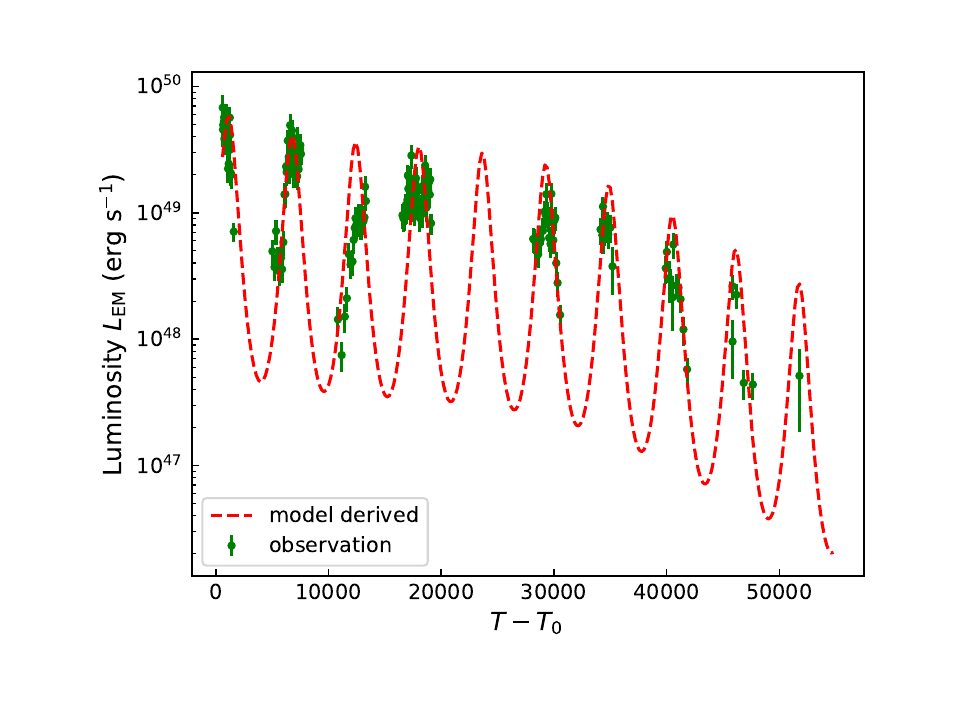}
\caption{A demonstration for QPO lightcurve. The observed lightcurve is shown in green scatter, while the lightcurve predicted by the precessional BH jet is represented by the red dashed line.}
\label{fig:light_curve}
\end{figure}

\begin{figure}[ht]
\centering
\includegraphics[angle=0,scale=0.6]{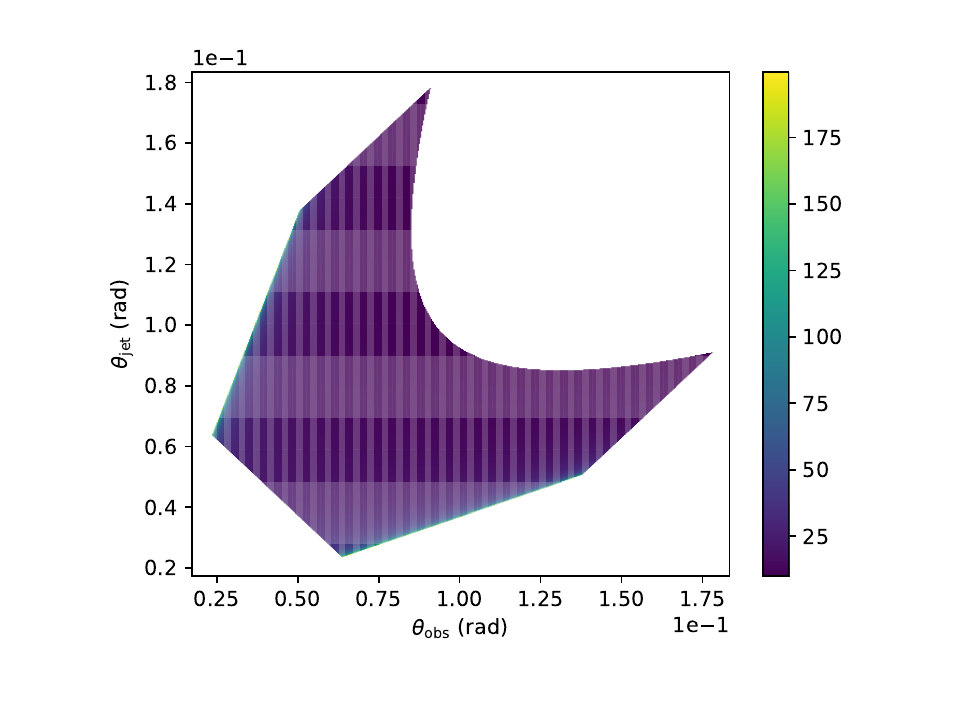}
\caption{
The pseudocolor plot illustrates values within the exclusive range of $10<\Gamma<200$. Intriguingly, in the majority of scenarios, the values of $\Gamma$ fall between 10 and 40, implying that the model favors a lower $\Gamma$. Additionally, the plot reveals a symmetrical distribution of $\theta_{\rm jet}$ and $\theta_{\rm obs}$\cite{}, with the viable range spanning from 0.02 rad to 0.18 rad.} 
\label{fig:Pseudocolor}
\end{figure}

The energetic gamma-ray burst known as GRB\,050904 was detected by the {\em Swift}/BAT instrument at 01:51:44 on September 4, 2005 (UT) \citep[][hereafter $T_0$]{2005GCN..3910....1C}.
The duration of its prompt emission was measured to be $T_{\rm 90} = 225\pm 10$ seconds \citep{2005GCN..3938....1S}.
\textit{Swift}/XRT instrument started to observe its X-ray emission since $T_0$+161 seconds.
The X-ray lightcurve, as observed, exhibits a total of nine flares from the end of the rapid decay phase (approximately 600 seconds) to 55,000 seconds,
and the contour line of the lightcurve is characteristic by a plateau followed by a sharp decay (with a decay index of $\alpha \sim -6$) \citep{2009MNRAS.397.1177E}.
A speculative period for successive flares is approximately 5600 seconds.
The confirmation of GRB\,050904 as the first GRB with a spectroscopic redshift larger than 6 was provided by subsequent optical observations \citep[z= 6.29;][]{2006Natur.440..184K}.
The isotropic X-ray luminosity of GRB\,050904 is derived with $L_{\rm iso,X}(t) = 4\pi D^2_L f_X(t) k(z)$, where $D_L$ and $f_X(t)$ are the luminosity distance and the observed X-ray flux, respectively. The cosmological $k$-correction factor has $k(z) = (1+z)^{\gamma-2}$ \citep{2001AJ....121.2879B,2019ApJ...886....5S}, where the average photon index of GRB\,050904 has $\gamma = 1.8 $\footnote{https://www.swift.ac.uk/index.php}.
The investigation of the jet opening angle for GRB050904 has been conducted extensively in previous studies \citep{2005A&A...443L...1T,2006ApJ...636L..69W,2006ApJ...646L..99F,2007ApJ...668.1083G}, suggesting a potential range of values from $3^{\circ}$ to $15^{\circ}$.
In this study, we adopt a moderate value of $\theta_j =10^{\circ}$, which corresponds to a beaming factor $f_b = 1-{\rm cos}\theta_j = 0.015$, assuming that the flares share the same jet opening angle as the prompt phase.
Considering a small Lorentz factor required in the below analysis, the energy fraction into XRT windows $\eta_{\rm X} = 0.1$ \citep{2019ApJ...878...62X} is
adopted to calculate the afterglow luminosity $L_{\rm obs} = L_{\rm iso,X} \times f_b/\eta_X$.
The resulting lightcurve is presented in Figure \ref{fig:light_curve}.
The observed flares are interpreted as a consequence of a precessing jet.
A numerical fitting process based on radiation from a precessional BZ jet is carried out subsequently.
Equation (\ref{eq:LEM}) demonstrates that the amplitude magnitude of the observed pulses is directly influenced by three parameters: luminosity $L_0$, Lorentz factor $\Gamma$, and angle $\psi$.
While searching for the parameter distribution, we fix certain parameters based on observations to resolve the degeneracy of the above three parameters.
The amplitude of pulses spans a range of two orders of magnitude, as depicted in Figure \ref{fig:light_curve}, allowing us to build an equation like below
\begin{eqnarray}
\left( \frac{1-\beta {\rm cos}(\psi_{\rm min})}{1-\beta {\rm cos}(\psi_{\rm max})}  \right)^3 \approx \frac{1}{100},
\label{eq:spn}
\end{eqnarray}
where $\beta < 1$ requires that
\begin{eqnarray}
 \frac{1-0.01^{1/3}}{{\rm cos}(\psi_{\rm min}) -0.01^{1/3} {\rm cos}(\psi_{\rm max})} <1.
\label{eq:betalesser1}
\end{eqnarray}
As the emission from a precessing jet can intermittently move in and out of the observer's line of sight, one expects that $\psi_{\rm min} = |\theta_{\rm jet}-\theta_{\rm obs} |\leq \theta_j/2 $ and $ \psi_{\rm max}= \theta_{\rm jet} + \theta_{\rm obs} > \theta_j/2 $.
In this scenario, a pseudocolor plot (Figure \ref{fig:Pseudocolor}) is constructed to visualize the range of $\theta_{\rm jet}$ and $ \theta_{\rm obs}$,
where the color in the plot represents the corresponding values of $\Gamma$.
The plot is configured with $10<\Gamma <200$, but the majority of values fall within the range of 10 to 40.
The given pseudocolor plot reveals a symmetrical distribution of $\theta_{\rm jet}$ and $ \theta_{\rm obs}$, with a viable range spanning from 0.02 rad to 0.18 rad.
\begin{deluxetable*}{ccccccccc}
\tablenum{1}
\tablecaption{Fitting result with the precessing jet model.}
\tablewidth{0pt}
\tablehead{
\colhead{$\theta_{\rm jet} $ (rad)} & \colhead{$\theta_{\rm obs} $ (rad)} & \colhead{$\Gamma$}& \colhead{${\rm log} L_0$}& \colhead{$P$ (s)}& \colhead{$C$}& \colhead{$\alpha_1$}& \colhead{$\alpha_2$}& \colhead{${\rm log}t_b$}
}
\startdata
0.04 & 0.07 & 21.47 &$49.61^{+0.14}_{-0.14}$ &  $5621^{+63}_{-72}$ &  $4.99^{+0.25}_{-0.27} $& $0.20^{+0.13}_{-0.12}$ & $6.39^{+1.07}_{-0.96}$ & $4.56^{+0.10}_{-0.07}$\\
0.06 & 0.11 & 14.56 &  $49.72^{+0.14}_{-0.14}$ &  $5620^{+64}_{-71}$ &  $4.99^{+0.26}_{-0.27} $& $0.20^{+0.13}_{-0.12}$ & $6.37^{+1.07}_{-0.96}$ & $4.56^{+0.10}_{-0.07}$\\
0.08 & 0.15 & 11.06  & $49.77^{+0.14}_{-0.14}$ &  $5622^{+64}_{-71}$ &  $4.99^{+0.26}_{-0.27} $& $0.20^{+0.13}_{-0.12}$ & $6.37^{+1.07}_{-0.95}$ & $4.56^{+0.10}_{-0.07}$\\
\enddata
\tablecomments{Based on the Figure \ref{fig:Pseudocolor}, we select three groups of $\theta_{\rm jet}$ and $\theta_{\rm obs}$ to test their impact on other parameters, and the $\Gamma$ is determinate by Equation (\ref{eq:spn}).}
\label{tab:three_pa}
\end{deluxetable*}

\begin{figure}[ht]
\centering
\includegraphics[angle=0,scale=0.4]{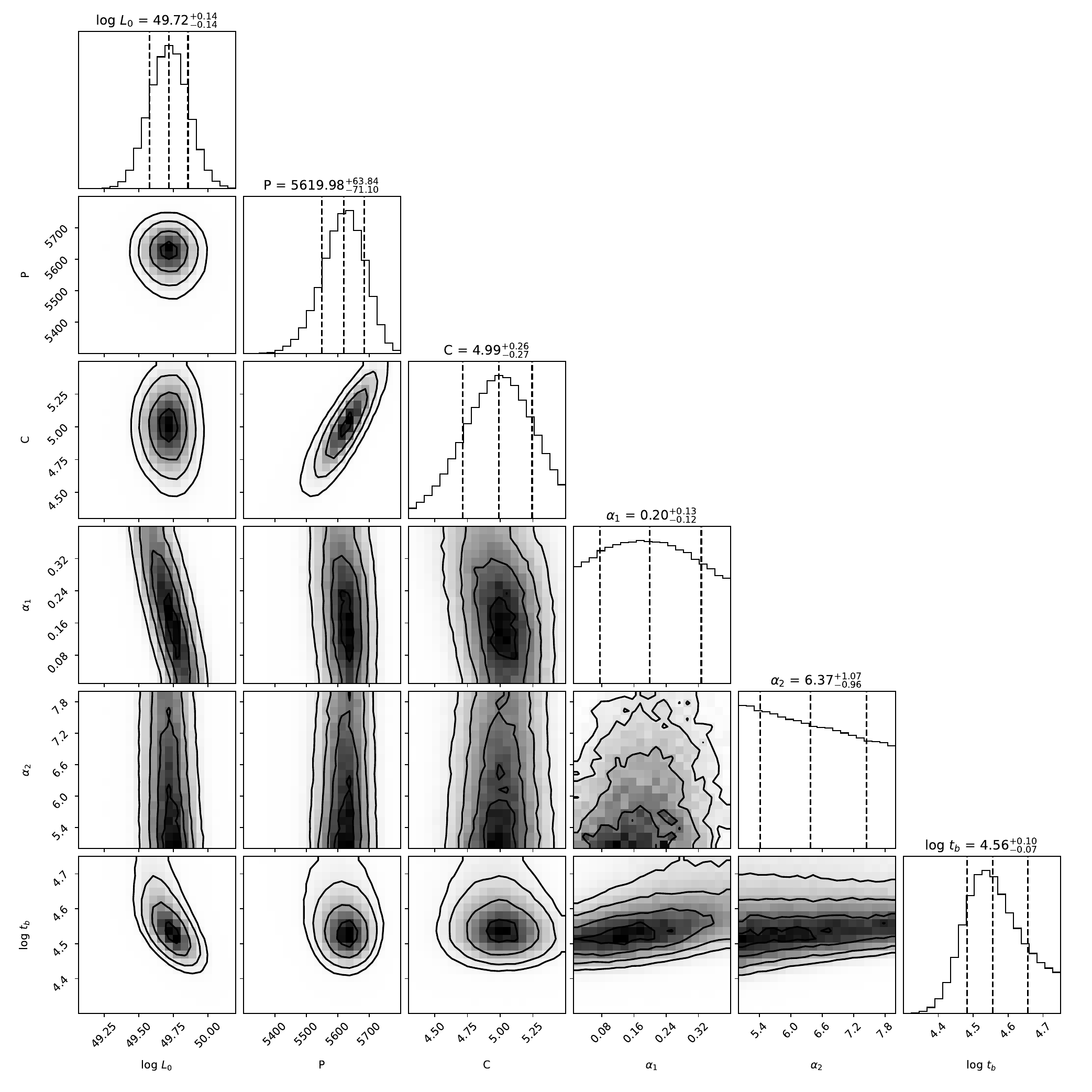}
\caption{The contour plot for the posterior distributions of parameters, corresponds to the scenario where $\theta_{\rm jet} = 0.06$ rad, $\theta_{\rm obs} = 0.11$ rad, and $\Gamma = 14.56$.}
\label{fig:Ps_space}
\end{figure}

Assuming $\theta_{\rm jet}<\theta_{\rm obs} $, we explore the parameters distribution by considering three groups of $\theta_{\rm jet}$, $\theta_{\rm obs}$, and $\Gamma $, as shown in Table \ref{tab:three_pa}.
We employed the MCMC method to explore the distribution of the parameters, and the results show good convergence with respect to the precession period.
The median values of the parameters, along with their 1 $\sigma$ uncertainty, have been tabulated in Table \ref{tab:three_pa}.
The results from the three different scenarios indicate that these three parameters have a negligible impact on the period of the precessing jet.
The corner plot in Figure \ref{fig:Ps_space} corresponds to the scenario where $\theta_{\rm jet} = 0.06$ rad, $\theta_{\rm obs} = 0.11$ rad, and $\Gamma = 14.56$. The resulting lightcurve is also shown in Figure \ref{fig:light_curve}.
The results of the numerical fitting indicate that the period of the precessing jet and the characteristic timescale in the rest frame are $P = 5620/(1+z) = 771 ~\rm s$ and $t_b = 10^{4.56}/(1+z) = 4980 ~\rm s$, respectively.
Assuming a dimensionless spin parameter $a \sim 1$ and a central BH mass of $M_{\rm BH} = 3 {\rm M_\odot}$, the equivalent radius is estimated to be $r =1.13 \times10^8 ~\rm cm$.
The derived characteristic luminosity suggests that the magnetic field at the BH horizon before it diffuses out, is as high as $B_{\rm H,0} = 9.49\times 10^{14}$ G.

\section{The Periodicity of observed X-ray flares}
\label{sect:4}
A potential periodicity in the detected flares was noticed by \cite{2006ApJ...637L..69W}, but no significant QPO signals were identified in their analysis.
In Section \ref{sect:3}, the result of the model fitting indicates the existence of the period of about 5600 s in the observer frame.
Therefore, it is valuable to investigate the possibility of identifying a periodic signal through a time series analysis of the observed data.
In this section, two methods are employed to test for periodicity.

\subsection{Preprocessing of observed lightcurve}

Section \ref{sect:Obpj} indicates that the observed periodic component is amplified with a power factor,
resulting from the Doppler effect, i.e., $L_{\rm EM} \propto F \propto D^3 ({\rm cos}(\psi(t)))$.
The least-squares method, however, is based on the linear sine function $y = a{\rm cos}(\omega) + b{\rm sin}(\omega) + const$.
Furthermore, the intrinsic luminosity follows a broken-power-law described trend.
Therefore, prior to testing for periodicity, the observed luminosity $L_{\rm EM}$ is transformed into the logarithmic form.
The intrinsic mean luminosity has
\begin{eqnarray}
L_{\rm mean} = D_{\rm mean}^3 L_{\rm EM}(\psi = 0),
\end{eqnarray}
where $D_{\rm mean}= \frac{1}{2}(D(\psi_{\rm max})+D(\psi_{\rm min}))$.
Therefore the lightcurve transforms to
\begin{eqnarray}
L_T = {\rm log}L_{\rm EM} - {\rm log}L_{\rm mean}.
\end{eqnarray}
Error propagation gives that $\sigma_T = \frac{1}{L_{\rm EM} \times {\rm ln}10} \sigma_{\rm EM}$,
where $\sigma_{\rm EM}$ and $\sigma_T$ are the error of $L_{\rm EM} $ and $L_T$, respectively.
Following Section \ref{sect:3}, we adopt that $\theta_{\rm jet} = 0.06 $ rad, $\theta_{\rm obs} =0.11$ rad, and $\Gamma = 14.56$,
the derived $L_T$ is displayed in figure \ref{fig:QPO_test} (a).

\subsection{Methods and test results}
\label{sec:test}
The method of least-squares was proposed to extract periods from inhomogeneous astronomical lightcurves by \cite{1976Ap&SS..39..447L}, its reliability tests were contributed by \cite{1982ApJ...263..835S}, this method is now known as Lomb-Scargle periodogram (LSP).
\cite{2009A&A...496..577Z} gave a generalized Lomb-Scargle periodogram (GLS) that takes errors into account and adds an offset constant to the sinusoidal function.
The wavelet transform overcomes the limitations of the Fourier transform through mathematical advancements.
\cite{1996AJ....112.1709F} developed the weighted wavelet Z-transform (WWZ) to detect pseudo-periodic signals in irregular time series.
\cite{2017eaydin.179..433A} developed a Python program for implementing the WWZ method.
The false alarm levels given by various methods are always based on statistics. However, the power-law component, red noise,
always raises the low-frequency port to a lower false alarm possibility (FAP).
\cite{2005A&A...431..391V} introduced a method for estimating significance in the context of the background noise following a chi-square distribution.
Furthermore, the background of the periodogram always exhibits two-smooth-broken-power-law behavior, which complicates the estimation of background noise.
By fitting a first-order autoregressive (AR1) procedure, \cite{2002CG.....28..421S} develop a Fortran program (REDFIT\footnote{http://www.geo.uni-bremen.de/~mschulz/}).

\begin{figure}[ht]
\gridline{\fig{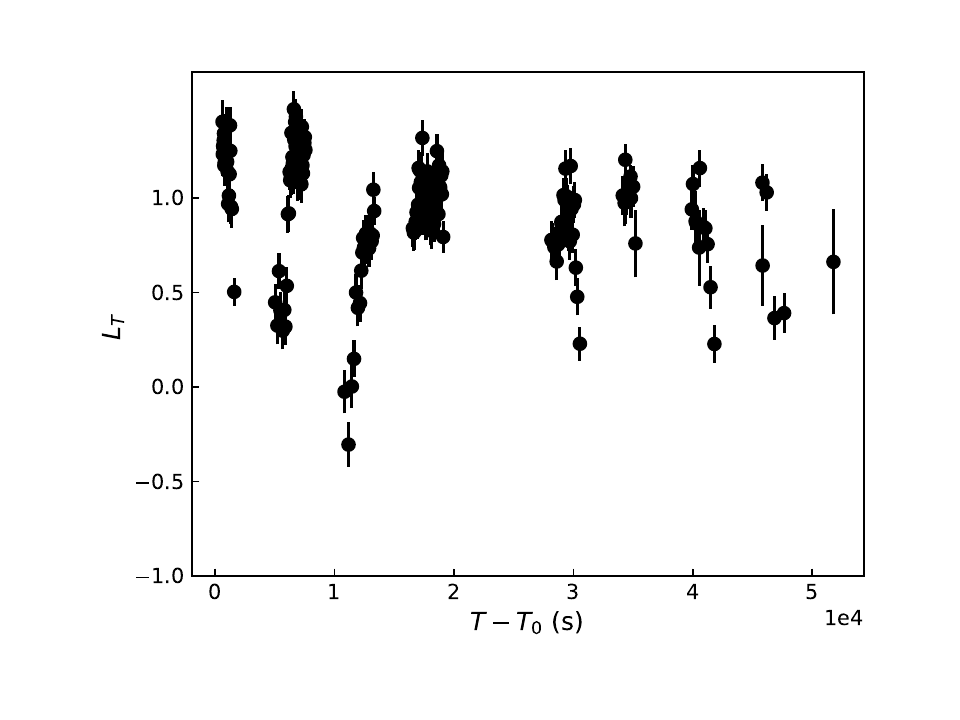}{0.45\textwidth}{(a)}}
\gridline{\fig{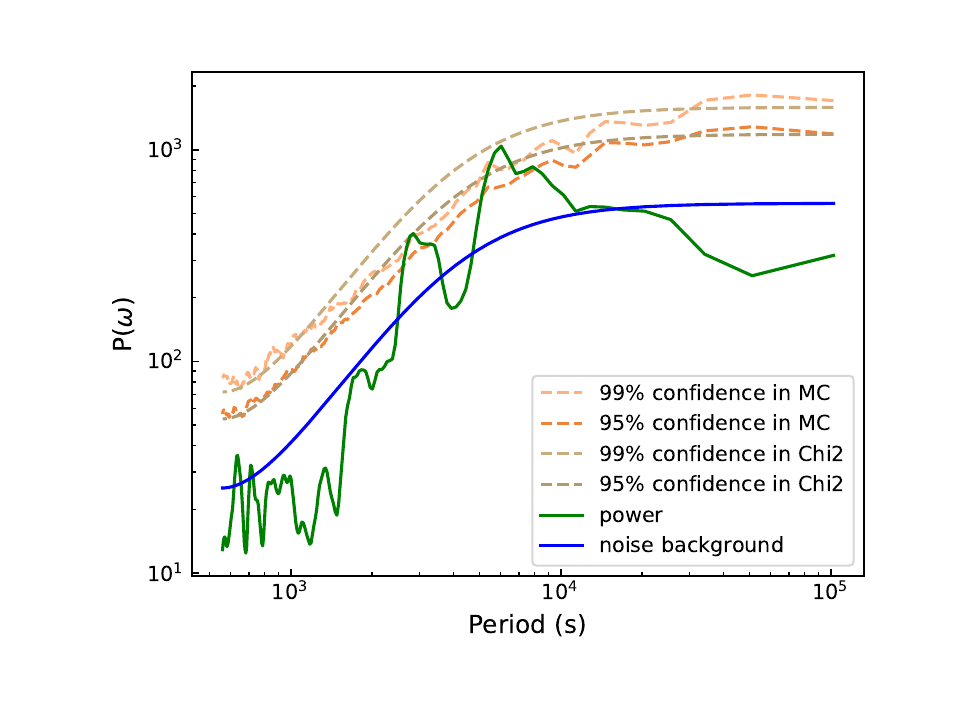}{0.45\textwidth}{(b)}
          \fig{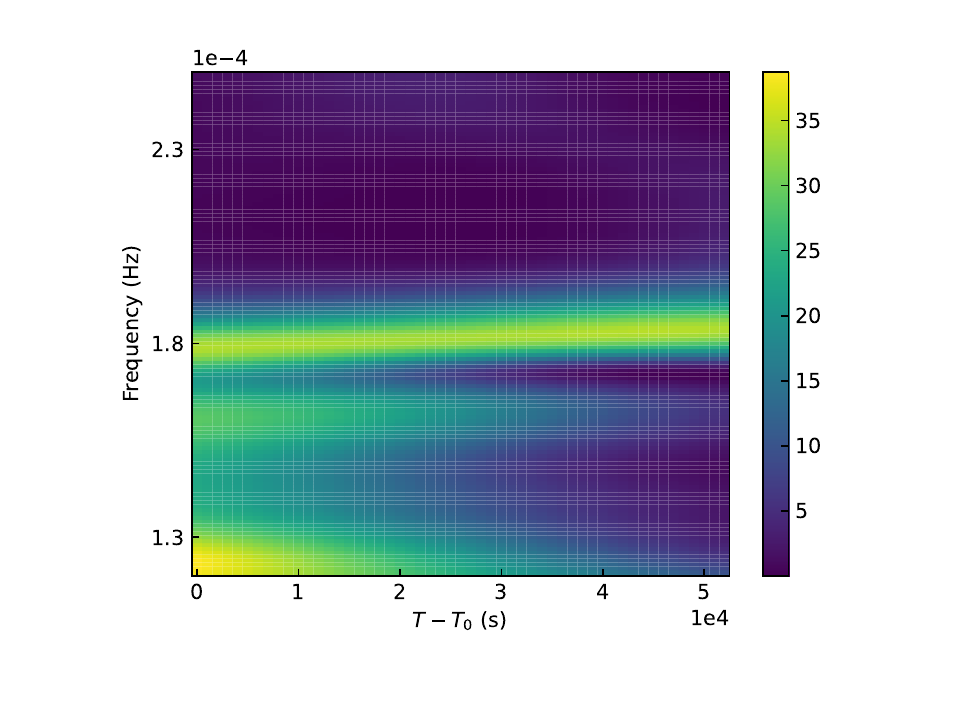}{0.45\textwidth}{(c)}}
\caption{(a) The derived lightcurve of $L_T$.
(b) The periodogram generated by REDFIT. The significance levels of 95\% and 99\% are performed in the dashed line.
(c) The periodogram generated by WWZ. A clear periodic signal local at $f = 1.8\times 10^{-4}~\rm Hz$.}
\label{fig:QPO_test}
\end{figure}

Panels (b) and (c) in Figure \ref{fig:QPO_test} represent the REDFIT-based periodogram and the WWZ-based wavelet transform, respectively.
REDFIT program offers confidence estimates for the corresponding LSP.
The confidence based on the Chi-square distribution of the spectrum is more stringent, with the signal at $P \sim 5600 \rm s$ having a confidence level of $Pr  > 95\% $, while the Monte Carlo ensemble provides a more flexible limit of $ Pr > 99\%$.
The wavelet transform reveals a robust and periodic signal at a frequency of approximately $f \sim 1.8 \times 10^{-4}~\rm Hz$ (corresponding to a period of $P \sim 5600 ~\rm s$).

\section{Conclusions and Discussions}
\label{sect:5}
Unlike a hairless BH, a magnetar possesses rich physical properties such as rapid rotation, high magnetic field, glitches, and magnetosphere, making it more conducive to interpreting various astrophysical phenomena.
AGN-powered jets, however, narrate a story of the interaction between a BH and its disk, facilitated by the presence of the magnetic field.
The origin of plateaus in GRB lightcurves is still a subject of debate.
A plateau, followed by either a shallow or steep decay, can be attributed to either internal or external origins.
The internal plateau, however, is usually attributed to a short-lived, supermassive magnetar.
A BH-based model, by taking the evolution of the magnetic field at the BH horizon into account, can also replicate a lightcurve featuring an internal plateau.
In this study, we propose that a precessional BZ jet would imprint a QPO signature on the internal plateau and the subsequent sharp decay.
Such lightcurves cannot be readily interpreted by the scenario of a short-lived magnetar, as it would have collapsed at the break time.
This model successfully reproduces a peculiar X-ray afterglow of GRB\,050904, which consists of nine flares, by assuming that the observed flares are intrinsic in nature.
One suggests that GRB\,050904 originates from a BH central engine, which powers a precessing jet with a period of 771 seconds.

Chromatic X-ray plateaus and optical afterglows have been extensively observed, such as GRB\,070110 \citep{2007ApJ...665..599T}. In GRB\,050904, the observed optical and near-infrared afterglow can be well explained by the standard afterglow model \citep[e.g.,][]{2006ApJ...636L..69W,2007ApJ...668.1083G}, while the X-ray flares attributed to internal dissipation of precessing jet. This fact implies a connection between X-ray flares and plateaus, in the afterglow of GRB\,050904.
The results of the periodicity test indicate that the confidence level for the periodicity of GRB\,050904 might not exceed $99\%$, a limitation largely attributable to the observational qualities.
GRB\,050904 is not a singular example presenting a potential QPO in the extensive X-ray afterglows collected by Swift/XRT, for example, GRB\,130925A is another source to exhibit possible QPO behavior \citep{2014ApJ...781L..19H}. However, Earth's shielding limits the ability of the XRT to capture comprehensive lightcurves of these GRBs.
The synergy between XRT and the forthcoming Einstein Probe \citep[{\em EP};][]{2022hxga.book...86Y} and Space-based Multiband Astronomical Variable Object Monitor \citep[{\em SVOM};][]{2016arXiv161006892W} observations of the same GRB afterglow is anticipated to yield a holistic lightcurve of these GRB afterglows.
Some researchers have suggested that the prompt emission possesses potential periodicity on time scales of seconds or less \citep{2022arXiv220507670Z,2022arXiv220502186X,2022ApJ...935..179W,2023RAA....23k5023Z}. Then whether these similar short-period activities exist in the X-ray plateau is very important for understanding the central engine of GRBs.
However, the sensitivity of XRT is insufficient for adequately capturing photons to discern these short-period activities.
NewAthena \citep[see also Athena: Advanced Telescope for High-ENergy Astrophysics;][]{2015JPhCS.610a2008B} may show extraordinary talent in the research of the periodicity in X-ray afterglows. NewAthena/WFI (Wide Field Imager) detection capability surpasses the XRT by more than an order of magnitude, which will enable a detailed examination of the temporal evolution of X-ray afterglows. Furthermore, EP/FXT (Follow-up X-ray Telescope) also exhibits enhanced performance in the lower energy band (0.5 - 2 keV) compared to XRT.


\begin{acknowledgments}
We acknowledge the use of the public data from the {\em Swift} data archive and the UK Swift Science Data Center.
We gratefully thank the anonymous referee for helpful comments to improve this paper. We thank Dr. Xiang Li for useful discussion. This  work was supported by NSFC (No. 12073080, 11933010, 11921003, 12303050) and by the Chinese Academy of Sciences via the Key Research Program of Frontier Sciences (No. QYZDJ-SSW-SYS024).
\end{acknowledgments}


\software{emcee \citep{2013PASP..125..306F},
REDFIT \citep{2002CG.....28..421S}, GLS \citep{2009A&A...496..577Z},
 WWZ \citep{2017eaydin.179..433A} }



\end{document}